\documentclass[aip,jcp,amsmath,amssymb,reprint,citeautoscript]{revtex4-2}

\usepackage{graphicx}
\usepackage{dcolumn}
\usepackage{bm}
\usepackage[utf8]{inputenc}
\usepackage[T1]{fontenc}

\usepackage{physics}                    
\usepackage{amsfonts}  
\usepackage{amsthm}    
\usepackage{mathtools} 
\usepackage{placeins}
\usepackage{booktabs}                   
\usepackage{xcolor}                     
\usepackage[english]{babel}

\newcommand{\mi}{\mathrm{i}}            

\newcommand{\cplus}{\ensuremath{\mathcal{C}^+}}
\newcommand{\br}{\ensuremath{\mathbf{r}}}
\newcommand{\bx}{\ensuremath{\mathbf{x}}}
\newcommand{\gKS}{\ensuremath{\mathrm{gKS}}}

\DeclareUnicodeCharacter{0308}{\"{}}

\graphicspath{ {./figures/} }

\begin{document}


\title{$\ensuremath{GW}$ reduced density matrix from iterated linearized Dyson equation}

\author{Fabien Bruneval}
 \email{fabien.bruneval@cea.fr}
 \affiliation{Universit\' e Paris-Saclay, CEA, Service de recherche en Corrosion et Comportement des Matériaux, SRMP, 91191 Gif-sur-Yvette, France}

\author{Erik Verzijl}
 \email{h.verzijl@vu.nl}
\affiliation{Department of Chemistry and Pharmaceutical Sciences, Vrije Universiteit Amsterdam, De Boelelaan 1105, 1081 HV Amsterdam, The Netherlands}%
\author{Arno Förster}
\email{a.t.l.foerster@vu.nl}
\affiliation{Department of Chemistry and Pharmaceutical Sciences, Vrije Universiteit Amsterdam, De Boelelaan 1105, 1081 HV Amsterdam, The Netherlands}%
\author{Mauricio Rodriguez-Mayorga}
 \email{marm3.14@gmail.com}
\affiliation{Laboratoire de Chimie et Physique Quantiques (UMR 5626), Universit\'e de Toulouse, CNRS, Toulouse, France}

\date{\today}

\begin{abstract}
Iterating the Dyson equation with the static part of the self-energy
leads to a concise and possibly improved expression of the one-body reduced density matrix
from any self-energy approximation.
Here we apply the procedure to Hedin's $GW$ approximation. 
The non-iterated $GW$ based density matrix was already known to yield accurate density matrices for molecular systems.
We show that the Dyson-equation-based procedure is equivalent to the so-called variational Z-vector approach applied
to the Random-Phase approximation energy functional,
but only in the case of a Hartree-Fock mean-field starting point.
When a generalized Kohn-Sham scheme is employed instead, the two approaches differ.
By comparing the density matrix for a benchmark set of 34 small molecules to coupled-cluster reference values,
we conclude that the iterated Dyson equation indeed produces improved density matrices for molecular systems.
Interestingly, we observe that the excitation rank of the reference coupled-cluster matters much and
that the inclusion of triple excitations (CCSDT) quantitatively changes the conclusions of the benchmark as compared to
single and double excitations coupled-cluster (CCSD).
\end{abstract}

\maketitle

\section{\label{sec:intro}Introduction}

When Hedin \cite{hedin_pr1965} wrote down in 1965 his celebrated system
of functional equations that solve, in principle, the many-electron problem,
the equations were meant to be solved self-consistently.
The Green's function $G$, the self-energy $\Sigma$, the screened Coulomb interaction $W$, etc.
were all supposed to be obtained via a self-consistent procedure.
This can be seen by the unique first-order in $W$ contribution (or Feynman diagram) that is considered.
Using non-self-consistent quantities would give rise to additional first-order contributions,
such as a correction to the Hartree potential.

Even considering the lowest-order approximation to the self-energy in Hedin's equations, namely the $GW$ approximation,
practical implementation of self-consistency is challenging.
\cite{holm_prb1998,garcia_prb2001,caruso_prb2012,koval_prb2014,kutepov_prb2009,grumet_prb2018,
yeh_prb2022,wen_jctc2024}
It should be noted here that the use of self-consistent Green's function has many
advantages.
It formally allows one to derive conserving approximations from the Baym and Kadanoff derivative condition \cite{baym_pr1961, baym_pr1962}.
Such a derived approximation will conserve the number of electrons, momentum, angular momentum, etc.
It has been shown recently that self-consistent Green's function can give access to derivatives \cite{Pokhilko2021b},
to measure spin contamination \cite{Pokhilko2022}, and calculate natural occupations \cite{Pokhilko2023}.

Self-consistency is in principle simple and elegant.
However, for practical convenience, the majority of the $GW$ calculations are performed as a one-shot correction to 
a mean-field starting point: This is the standard $G_0W_0$ procedure
\cite{reining_wires2018, marie2024gw}.
When skipping the self-consistency for the Green's function, 
additional low-order diagrams have to appear.
They correspond to the static part of the self-energy,
which arises from changes in the Hartree and exchange potentials.
These diagrams are carefully taken into account in the conventional perturbation theory
based on the bare Coulomb interaction $v$ as introduced by Cederbaum and coworkers \cite{cederbaum_book1977,vonniessen_cpr1984}.
In perturbation theory at the third order (PT3 or GF3), static self-energy contributions are indeed present.
Quite the opposite, in the $GW$ community, these static diagrams are most often ignored.
However, their effect is substantial, and they generally improve the quality of the quasiparticle energies\cite{bruneval_fchem2021, tolle2026connection},
as shown with the $GW$ density matrix term $\gamma^{GW}$ evaluated by one of us a few years back
\cite{bruneval_prb2019,bruneval_jctc2019,bruneval_jctc2021}.
Based on the linearized Dyson equation, it was possible to provide a closed-form expression for
the $GW$ density matrix. This will be recapped in Section~\ref{sec:recap}.

But Cederbaum, Schirmer, and coworkers, working on elaborating the algebraic diagrammatic construction (ADC) formalism,
went one step beyond the one-shot linearized density matrix in iterating the 
Dyson equation with an updated self-energy for the static part solely
\cite{schirmer_pra1983,vonniessen_cpr1984,schirmer_jchemphys1989, Trofimov2005MolecularApproach}.
As we will show here, the density matrix obtained in this way corresponds to the relaxed density matrix.

In the present work, we apply the same idea to the $GW$ density matrix.
We will furthermore consider the possibility of using any generalized Kohn-Sham (gKS) starting point
as it is customary in the $GW$ community, \cite{reining_wires2018}
whereas Cederbaum, Schirmer, and coworkers always relied on a Hartree--Fock (HF) mean-field,
which nullifies the first-order terms by virtue of the Brillouin theorem \cite{helgaker_book}.
This theory will be developed in Section~\ref{sec:theory}.

Finally, the practical performance of the proposed expressions will be assessed
on a molecular benchmark of 34 small molecules in Section~\ref{sec:results},
where comparison to coupled-cluster reference results will be undertaken.

\section{\label{sec:recap} Quick recap about $GW$ density matrix}

\subsection*{Notations}

Hereafter we employ the Hartree atomic units where
$\hbar = m_e =  e = 1 / 4\pi \epsilon_0 = 1$.

The equations are formulated in the orthogonal molecular orbital (MO) basis obtained from a prior self-consistent 
gKS calculation.
We use the standard convention that indices
$p, q, m, n$ denote general MOs,
indices $i, j, k$ run over occupied MOs,
and $a, b, c$ over empty MOs.
We use spin-orbitals,
the corresponding spatial orbital expressions can be retrieved by adding spin factors.

In the following, we use both Coulomb integrals in 
the chemist's notation:
\begin{equation}
  ( p q | m n ) = \iint d \bx d \bx'
     \varphi^*_p(\bx)  \varphi_q(\bx) \frac{1}{|\br-\br'|}
     \varphi^*_m(\bx') \varphi_n(\bx')
\end{equation}
and antisymmetrized Coulomb integrals in 
physicist's notation:
\begin{equation}
  \langle  pm || qn \rangle = (p q | m n) - (p n | m q) .
\end{equation}

\subsection{Density matrix from a Green's function using the linearized Dyson equation}

The time-ordered single-particle Green's function $G(\br t , \br' t' )$ contains a great deal of information. \cite{fetter_book}
Besides the ionization and affinity energies and the total energy, it also gives access to the
electronic density and density matrix when evaluated in the equal time limit at $t' \to t^+$.

\begin{figure}
\includegraphics[width=0.95\columnwidth]{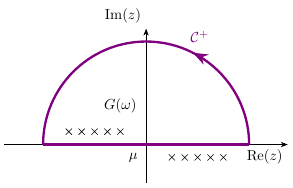}
\caption{\label{fig:contour}
Contour integration \cplus in the complex plane.
The poles of the time-ordered Green's function $G$ are marked as crosses.
The contour encloses the poles in the upper half plane that corresponds to occupied states.
The Fermi level $\mu$ splits the occupied poles (in the second quadrant)
and the virtual poles (in the fourth quadrant).
}
\end{figure}

For our purposes, it is more convenient to express this density matrix $\gamma$ in the MO basis and in the frequency domain:
\begin{equation}
  \label{eq:gamma_g}
  \gamma_{pq} = \frac{1}{2\pi \mi} \int_{\mathcal{C}^+} d \omega G_{pq}(\omega)
\end{equation}
where \cplus is a closed contour going from $-\infty$ to $+\infty$ and a half circle in the upper complex plane
as drawn in Fig.~\ref{fig:contour}. The contour can only be closed in the upper half-plane because of the equal-time limit.
Hereafter, we write $\gamma_{pq}^\gKS$ for the density matrix obtained from the gKS single-particle Green's function $G^\gKS$.

The self-energy $\Sigma(\omega)$ is defined as the dynamical
operator that connects a known
mean-field Green's function $G^\gKS$ to a fully interacting Green's function $G$ thanks to the Dyson equation.
Formally, one can always split the self-energy in a 
static part $\Sigma(\infty)$ and a dynamical part $M(\omega)$ (also named mass operator),
where we use the notations of Cederbaum and coworkers \cite{vonniessen_cpr1984}.
Mathematically, $\Sigma(\infty)$ is defined as the limit of the self-energy
when its argument $\omega$ goes to infinity.
Indeed in this limit, all the contributions having a pole vanish and 
only the static part survives.

Now analyzing the static self-energy $\Sigma(\infty)$,
the Feynman diagrams drawn by Cederbaum and coworkers \cite{vonniessen_cpr1984}
show that the only contributions to $\Sigma(\infty)$ are
the Hartree potential, the Fock exchange, and possibly
the KS potential if starting from a generalized Kohn-Sham Green's function
\cite{ren_prl2011}.

To proceed, we write the linearized Dyson equation:
\begin{multline}
  \label{eq:ldyson}
  G_{pq}(\omega) \approx G_{pp}^\gKS(\omega)  \delta_{pq}
    + G_{pp}^\gKS(\omega)  \\
      \cdot
     \big\{
      \langle p | \Sigma_x[\gamma^\gKS] - v_{xc}[\gamma^\gKS] | q \rangle \\
     + M_{pq}(\omega) 
     \big\}
      G_{qq}^\gKS(\omega) ,
\end{multline}
where we have used the fact that the gKS Green's function
\begin{equation}
G_{pq}^\gKS(\omega) = \frac{\delta_{pq}}{\omega - \epsilon_p + \mi \eta \cdot \mathrm{sign}(\epsilon_p - \mu)}
\end{equation}
is diagonal in the MO basis, and $\mu$ is the chemical potential introduced to distinguish the occupied and virtual states.

With this definition, the complete self-energy $\Sigma$, i.e. the term in between curly brackets in Eq.~(\ref{eq:ldyson}),
contains 
$v_{xc}$, the exchange-correlation potential that was used to generate the mean-field $G^\gKS$.
This exchange-correlation potential may or may not contain a fraction of non-local exchange depending
on the functional we chose.
In the specific case of HF, $v_{xc}$ reduces to $\Sigma_x$ and the self-energy $\Sigma(\omega)$
simply equals $M(\omega)$.

Combining Eqs.~(\ref{eq:gamma_g}) and~(\ref{eq:ldyson}), we obtain a working equation of the Green's function-derived density matrix:
\begin{equation}
  \label{eq:deltagamma}
  \Delta \gamma_{pq} = 
         \frac{1}{2\pi \mi} \int_{\cplus}
               d \omega
                  G_{pp}^\gKS(\omega) \Sigma_{pq}(\omega) G_{qq}^\gKS(\omega) ,
\end{equation}
where we have introduced a handy notation for the correction to the density matrix:
\begin{equation}
  \Delta \gamma_{pq} = \gamma_{pq} - \gamma_{pq}^\gKS .
\end{equation}

With this at hand, we are ready to specify the density matrix from the $GW$ approximation, $\gamma_{pq}^{GW}$.

\subsection{$GW$ approximation}

Here we briefly summarize the key steps that were used in previous derivations
\cite{bruneval_prb2019,bruneval_jctc2019,bruneval_jctc2021}.

Within $GW$ and with a one-shot evaluation,
the dynamic part of the self-energy is obtained via a convolution between $G^\gKS$ and $W_p = W - v$, the polarizable
part of $W$:
\begin{equation}
  \label{eq:gwp}
  M(\omega) = \frac{\mi}{2\pi} \int d \omega^\prime G(\omega + \omega^\prime) W_p(\omega^\prime) .
\end{equation}

This integral can be performed analytically to obtain
the $GW$ self-energy \cite{tiago_prb2006,vansetten_jctc2013,bruneval_cpc2016,marie2024gw}
\begin{equation}
  \label{eq:gwp_analytic}
  M_{pq}(\omega) = \sum_{ks} \frac{w^s_{kp} \cdot w^s_{kq}}{\omega - \epsilon_k + \Omega_s - \mi \eta}
                  +\sum_{cs} \frac{w^s_{cp} \cdot w^s_{cq}}{\omega - \epsilon_c - \Omega_s + \mi \eta} ,
\end{equation}
provided that $w^s_{pq}$ and $\Omega_s$ were obtained from the diagonalization of a Casida-like equation within the random-phase approximation
\begin{equation}
  \begin{pmatrix}
     {\bf A}  & {\bf B} \\
    -{\bf B} & -{\bf A}
  \end{pmatrix}
  \cdot
  \begin{pmatrix}
    {\bf X}^s \\
    {\bf Y}^s
  \end{pmatrix}
  = \Omega_s
  \begin{pmatrix}
    {\bf X}^s \\
    {\bf Y}^s
  \end{pmatrix} ,
\end{equation}
leading to  
\begin{equation}
   w_{kp}^s = \sum_{ ia} ( k p | i a )  \left( X_{ia}^s + Y_{ia}^s \right) .
\end{equation}
The expression of blocks ${\bf A}$ and ${\bf B}$ can be found, for instance, in Ref.~\onlinecite{bruneval_cpc2016}.

Then again performing the integral in Eq.~(\ref{eq:deltagamma}) analytically,
we obtain the expressions for the different blocks of the $GW$ density matrix:
\begin{subequations}
\begin{align}
  \label{eq:gwoo}
  \Delta \gamma^{GW}_{ij}
     = &- \sum_{sc} \frac{w^s_{ic}}{\epsilon_i - \epsilon_c -\Omega_s} 
              \cdot \frac{w^s_{jc}}{\epsilon_j - \epsilon_c -\Omega_s} \\
  \label{eq:gwvv}
  \Delta \gamma^{GW}_{ab}
     = & \phantom{-} \sum_{sk} \frac{w^s_{ak}}{\epsilon_a - \epsilon_k +\Omega_s} 
                         \cdot \frac{w^s_{bk}}{\epsilon_b - \epsilon_k +\Omega_s} \\
  \label{eq:gwov}
  \Delta \gamma^{GW}_{ib}
     = & 
       \frac{1}{\epsilon_i - \epsilon_b}
       \Bigg\{
      \sum_{sk} \frac{w^s_{ik} \cdot w^s_{bk}}
                     {\epsilon_b - \epsilon_k +\Omega_s} \nonumber\\
      & 
       \phantom{ \frac{1}{\epsilon_i - \epsilon_b} }
      + \sum_{sc} \frac{w^s_{ic} \cdot w^s_{bc}}
                     {\epsilon_i - \epsilon_c -\Omega_s}  
                      \nonumber\\
      & 
       \phantom{ \frac{1}{\epsilon_i - \epsilon_b} }
       + \langle i | \Sigma_x[\gamma^\gKS] - v_{xc}[\gamma^\gKS] | b \rangle
       \Bigg\}
\end{align}
\end{subequations}

This expression for the density matrix has been shown to yield much improved results
for ionization potentials \cite{bruneval_prb2019,bruneval_fchem2021,tolle2026connection}
or
for electronic densities \cite{bruneval_jctc2019,bruneval_jctc2021}.
Alternatively, the integral in Eq.~\eqref{eq:deltagamma} can also be evaluated numerically via integration over the imaginary frequency axis,\cite{Ramberger2019,denawi2023gw}
providing a practical route for its evaluation within low-scaling $GW$ implementations, avoiding the explicit computation of $w_{pq}^s$ and $\Omega_s$.

\section{\label{sec:theory} Density matrix from iterated Dyson equation}

\subsection{Linearized Dyson equation with updated density matrix}

Now we will follow Cederbaum's idea to update the density matrix (see Appendix~B of Ref.~\onlinecite{vonniessen_cpr1984})
and generalize it to any gKS mean-field.

In this case, the linearized Dyson equation reads
\begin{multline}
  \label{eq:ldyson_gks}
  G_{pq}(\omega) \approx G_{pq}^\gKS(\omega) \delta_{pq}
     + G_{pp}^\gKS(\omega)  \\
       \cdot
      \big\{ \langle p | v_H[\gamma] + \Sigma_x[\gamma] 
        - v_H[\gamma^\gKS] - v_{xc}[\gamma^\gKS] | q \rangle  \\
      + M_{pq}(\omega) 
         \big\}
         G_{qq}^\gKS(\omega) .
\end{multline}
It is instructive to inspect the differences with the previous Dyson equation reported in Eq.~(\ref{eq:ldyson}).
In Eq.~(\ref{eq:ldyson_gks}) we have updated the functional dependence of the Hartree and exchange potentials
on the final density matrix $\gamma$, which is at this stage unknown.
Since the density matrix is changed, the Hartree potential that was not present in Eq.~(\ref{eq:ldyson}) now appears explicitly in 
Eq.~(\ref{eq:ldyson_gks}).
Note that inserting $\gamma = \gamma^\gKS$ in Eq.~(\ref{eq:ldyson_gks}) brings us back to Eq.~(\ref{eq:ldyson}):
the Hartree potentials cancel out and only $\Sigma_x[\gamma^\gKS]- v_{xc}[\gamma^\gKS]$ survives.

Let us identify the static part of the self-energy $\Sigma(\infty)$:
\begin{equation}
  \label{eq:sigmainftya}
  \Sigma_{pq}(\infty) 
      = \langle p | v_H[\gamma] + \Sigma_x[\gamma]  - v_H[\gamma^\gKS] - v_{xc}[\gamma^\gKS] | q \rangle  .
\end{equation}
Since the Hartree and exchange terms are linear with respect to their functional dependence on the density matrix, we can exploit this linearity to rewrite Eq.~(\ref{eq:sigmainftya}) as a function of the difference $\Delta \gamma$:
\begin{multline}
  \label{eq:sigmainftyb}
  \Sigma(\infty) 
      = \langle p | v_H[\Delta \gamma] + \Sigma_x[\Delta \gamma] | q \rangle \\
        + \langle p | \Sigma_x[\gamma^\gKS] - v_{xc}[\gamma^\gKS] | q \rangle .
\end{multline}

To proceed further, we specify the expression of Hartree and Fock potentials in
terms of antisymmetrized integrals:
\begin{multline}
  \label{eq:sigmainfty}
  \Sigma_{pq}(\infty) 
     = \sum_{mn} \langle  pm || qn \rangle \Delta \gamma_{mn} \\
        + \langle p | \Sigma_x[\gamma^\gKS] - v_{xc}[\gamma^\gKS] | q \rangle .
\end{multline}

Now we inject this final expression for the self-energy $\Sigma(\infty) + M(\omega)$
in the equation for $\Delta \gamma$ from Eq.~(\ref{eq:deltagamma}):
\begin{align}
  \label{eq:idgw_int}
  \Delta \gamma_{pq}^{\textrm{id}GW} =&
         \frac{1}{2\pi \mi} \int_{\cplus}
               d \omega
                  G_{pp}^\gKS(\omega) \Sigma_{pq}(\infty) G_{qq}^\gKS(\omega) \nonumber \\
                 & + 
         \frac{1}{2\pi \mi} \int_{\cplus}
               d \omega
                  G_{pp}^\gKS(\omega) M_{pq}(\omega) G_{qq}^\gKS(\omega) \;,
\end{align}
where we have introduced the superscript ``id$GW$'' for iterated-Dyson $GW$.

Because $\Sigma(\infty)$ is static, the frequency integral in the first term of Eq.~(\ref{eq:idgw_int})
is non-zero only when the index pair ($p$, $q$) is either occupied-virtual or virtual-occupied.
As a consequence, the occupied-occupied and virtual-virtual blocks are just $\int G^\gKS M G^\gKS$
as in the case of the $GW$ density matrix:
\begin{subequations}
\begin{align}
  \label{eq:idgwoo}
  \Delta \gamma_{ij}^{\textrm{id}GW} & = \Delta \gamma_{ij}^{GW} \\
  \label{eq:idgwvv}
  \Delta \gamma_{ab}^{\textrm{id}GW} & = \Delta \gamma_{ab}^{GW} \;,
\end{align}
\end{subequations}
and the iterated Dyson procedure affects only the occupied-virtual block. Mathematically, this reflects the fact that orbital relaxation effects are unitary transformations of the underlying mean field, and by Thouless' theorem, these are exclusively generated by particle-hole excitations and de-excitations.\cite{thouless1960stability} 

These equalities of the occupied-occupied and virtual-virtual blocks
in Eq.~(\ref{eq:idgwoo}, \ref{eq:idgwvv}) certify that the traces of
$\Delta \gamma^{\textrm{id}GW}$ and of
$\Delta \gamma^{GW}$ are equal.
Hence $\Delta \gamma^{\textrm{id}GW}$ inherits the
important electron number conservation property from
$\Delta \gamma^{GW}$ as shown in Ref.~\onlinecite{bruneval_jctc2021}.

There are two contributions in the self-energy
from Eq.~(\ref{eq:sigmainfty}) in addition to the one originated from the dynamical part $M(\omega)$:
\begin{align}
  \Delta \gamma_{ia}^{\textrm{id}GW} =&
       \frac{1}{\epsilon_i - \epsilon_a} 
              \sum_{mn} \langle  im || an \rangle \Delta \gamma_{mn}^{\textrm{id}GW} \nonumber\\
      &+\frac{1}{\epsilon_i - \epsilon_a}
  \langle i | \Sigma_x[\gamma^\gKS] - v_{xc}[\gamma^\gKS] | a \rangle \nonumber\\
                 & + 
         \frac{1}{2\pi \mi} \int_{\cplus}
               d \omega
                  G_{ii}^\gKS(\omega) M_{ia}(\omega) G_{aa}^\gKS(\omega)
\end{align}
 The two latter terms are independent from $\Delta \gamma_{mn}^{\textrm{id}GW}$ and
 combine to form $\Delta \gamma_{ia}^{GW}$ as in Eq.~(\ref{eq:gwov}).
 At the end of the day, we obtain
\begin{equation}
  \Delta \gamma_{ia}^{\textrm{id}GW}  = \frac{1}{\epsilon_i - \epsilon_a}
          \sum_{mn} \langle  im || an \rangle \Delta \gamma_{mn}^{\textrm{id}GW} 
                  + \Delta \gamma_{ia}^{GW} 
\end{equation}
In order to get a closed equation of $\Delta \gamma_{ia}^{\textrm{id}GW}$, we just need
to specify $\Delta \gamma_{mn}^{\textrm{id}GW}$ in the right-hand side of the previous equation.
We use Eqs.~(\ref{eq:idgwoo}, \ref{eq:idgwvv}) and the symmetry of density-matrices
$\Delta \gamma_{bj}^{\textrm{id}GW} = \Delta \gamma_{jb}^{\textrm{id}GW}$ and finally write
\begin{align}
  \label{eq:idgwov}
  \Delta \gamma_{ia}^{\textrm{id}GW}  = &
\frac{1}{\epsilon_i - \epsilon_a}
          \sum_{jb} \left( \langle  ij || ab \rangle 
                        +  \langle  ib || aj \rangle
                    \right)
                      \Delta \gamma_{jb}^{\textrm{id}GW} \nonumber \\
        & + \frac{1}{\epsilon_i - \epsilon_a}
            \sum_{kl} \langle  ik || al \rangle \Delta \gamma_{kl}^{GW}  \nonumber \\
        & + \frac{1}{\epsilon_i - \epsilon_a}
            \sum_{bc} \langle  ib || ac \rangle \Delta \gamma_{bc}^{GW}  \nonumber \\
        & + \Delta \gamma_{ia}^{GW} 
\end{align}

This equation for $\Delta \gamma_{ia}^{\textrm{id}GW}$ can be solved by iterating, by solution of the linear system, or by numerical integration as we show in the following subsections.

\subsection{Linear system}\label{sec:LinSys}

Now we show how Eq.~(\ref{eq:idgwov}) can be solved by solving a linear system of equations.
The unknown is $\Delta \gamma_{ia}^{\textrm{id}GW}$.
Let us multiply Eq.~(\ref{eq:idgwov}) by $(\epsilon_a - \epsilon_i)$ and swap terms:
\begin{multline}
  \sum_{jb}
\bigg[ 
(\epsilon_a - \epsilon_i) \delta_{ij} \delta_{ab} 
  + \langle  ij || ab \rangle 
      +  \langle  ib || aj \rangle
\bigg] 
\cdot 
\Delta \gamma_{jb}^{\textrm{id}GW}   \\
       =
          -\sum_{kl} \langle  ik || al \rangle \Delta \gamma_{kl}^{GW}   
          -\sum_{bc} \langle  ib || ac \rangle \Delta \gamma_{bc}^{GW}   \\
          +(\epsilon_a - \epsilon_i)  \Delta \gamma_{ia}^{GW} .
\end{multline}

This equation takes the form of a linear system, $\boldsymbol{\mathcal{A}} \cdot \boldsymbol{\mathcal{X}} = \boldsymbol{\mathcal{Y}}$,
where the matrix $\boldsymbol{\mathcal{A}}$ and the vectors $\boldsymbol{\mathcal{X}}$ and $\boldsymbol{\mathcal{Y}}$ have the following expressions:
\begin{subequations}
\begin{align}
  \label{eq:linear_system_a}
  \mathcal{A}_{ia, jb} = &  
(\epsilon_a - \epsilon_i) \delta_{ij} \delta_{ab} 
  + \langle  ij || ab \rangle 
      +  \langle  ib || aj \rangle \\
  \label{eq:linear_system_x}
  \mathcal{X}_{jb} = & \Delta \gamma_{jb}^{\textrm{id}GW}  \\
  \label{eq:linear_system_y}
  \mathcal{Y}_{ia} = & -\sum_{kl} \langle  ik || al \rangle \Delta \gamma_{kl}^{GW}   \nonumber  -\sum_{bc} \langle  ib || ac \rangle \Delta \gamma_{bc}^{GW}   \nonumber \\
           & +(\epsilon_a - \epsilon_i)  \Delta \gamma_{ia}^{GW} .
\end{align}
\end{subequations}

In Appendix~\ref{app:alternate}, we provide an alternate expression for the linear system 
that may be more convenient for implementation.

\subsection{Numerical integration}\label{sec:NumInt}

Alternatively, we can evaluate the integral in Eq.~\eqref{eq:deltagamma} numerically on an imaginary frequency grid.

Let us introduce the $N_k$ quadrature points $\omega_k$ with weights $\sigma_k$.
We can again identify three different contributions,
\begin{align}\label{eq:deltagammanumerically}
  \Delta \gamma_{pq}^{\textrm{id}GW} =&
       \frac{1}{2\pi}\sum_k^{N_k}\sigma_k
                  G_{pp}^\gKS(\mu + \mi \omega_k) \nonumber \\
             &   ~ \cdot 
             \bigl\langle p \bigl| 
                  v_H[\Delta\gamma^{\textrm{id}GW}]
                  + \Sigma_x[\Delta\gamma^{\textrm{id}GW}]
            \bigr| q \bigr\rangle 
                  \nonumber \\ 
             & ~ \cdot G_{qq}^\gKS(\mu + \mi \omega_k) \nonumber\\
      &+\frac{1}{2\pi}\sum_k^{N_k}\sigma_k
                  G_{pp}^\gKS(\mu + \mi \omega_k) \nonumber \\
                 & ~ \cdot 
                 \bigl\langle p \bigl| \Sigma_x[\gamma^\gKS]
                    - v_{xc}[\gamma^\gKS] 
                    \bigr| q \bigr\rangle \nonumber \\
                 & ~ \cdot G_{qq}^\gKS(\mu + \mi \omega_k) \nonumber\\
                 & + \frac{1}{2\pi}\sum_k^{N_k}\sigma_k 
                  G_{pp}^\gKS(\mu + \mi \omega_k)\nonumber \\
                & ~ \cdot M_{pq}(\mu + \mi \omega_k)
                  G_{qq}^\gKS(\mu + \mi \omega_k) \;.
\end{align}
Because the static part of the self-energy yields non-vanishing contributions only for occupied-virtual (and virtual-occupied) index pairs, the occupied-occupied and virtual-virtual blocks should converge to zero with increasing grid size. Consequently, the calculation simplifies to the numerical integration of $\int G^\gKS_{ii}\Sigma_{ia}(\infty)G^\gKS_{aa}$. 

In this numerical approach, $\Delta\gamma^{\textrm{id}GW}$ cannot be cast as a linear system. This requires a self-consistent approach, in which linear mixing, direct inversion of the iterative subspace,\cite{pulay_cpl1980} or quasi-Newton methods\cite{johnson1988modified} can be employed to stabilize and/or accelerate convergence. Since the latter two contributions in Eq.~\eqref{eq:deltagammanumerically} do not depend on $\Delta\gamma^{\textrm{id}GW}$, they provide the initial guess for the self-consistent loop that is used to solve for the contribution from the first term iteratively.

\subsection{Comparison with Z-vector technique relaxed density matrix}

In the context of gradient evaluation from correlated methods, Handy and Schaefer \cite{handy_jcp1984} introduced the so-called Z-vector technique.
The Z-vector technique produces the so-called relaxed density matrix based on the knowledge of the unrelaxed one. Here in our $GW$ context, the unrelaxed density matrix would be precisely $\Delta \gamma^{GW}$ as shown in
Eqs.~(\ref{eq:gwoo}-\ref{eq:gwov}).

Using the Z-vector technique, Rekkedal~\textit{et. al.} \cite{rekkedal2013communication} and Burow~\textit{et. al.} \cite{burow_jctc2014} proposed an expression for the relaxed density matrix 
corresponding to the RPA energy functional.
Comparing our linear system in Eqs.~(\ref{eq:linear_system_a}-\ref{eq:linear_system_y})
obtained from the iterated Dyson equation
with Eq.~(33) in Ref.~\onlinecite{burow_jctc2014} obtained from the variation of a Lagrangian, we find that the terms correspond one-by-one,
with the exception of the electronic Hessian (or orbital stability matrix).

The iterated Dyson expression for matrix $\boldsymbol{\mathcal{A}}$ in Eq.~(\ref{eq:linear_system_a}) corresponds to 
the HF electronic Hessian, whereas the Hessian in Eq.~(34) of Ref.~\onlinecite{burow_jctc2014} is the one of the gKS mean-field:
\begin{align}
  \mathcal{A}^\mathrm{gKS}_{ia, jb}
  = & 
(\epsilon_a - \epsilon_i) \delta_{ij} \delta_{ab} 
  +  2 (ia | jb ) \nonumber \\
  &
  -  \alpha \bigl[ (ib  | ja) + ( i j | b a) \bigr]
  + 2 \langle  i a | f_{xc} | j b \rangle \:,
\end{align}
where we have used the expression for a global hybrid
with $\alpha$ content of exact-exchange and
an exchange-correlation kernel $f_{xc}$.

This means that our expressions only match when applied on top of HF,
and in this case $\alpha=1$ and $f_{xc}=0$.
This also proves that the iterated Dyson density matrix is the analytical derivative of the RPA total energy for a HF starting point.

Let us confirm these conclusions in a practical example.
We consider the evaluation of the static dipole of the carbon monoxide dimer with two methods.
Either we directly use the density matrix, 
\begin{equation}
  \mathbf{D} = - \sum_{mn} \gamma_{mn} \langle m | \mathbf{r} | n \rangle + \sum_{\mathbf{R}_A} Z_A \mathbf{R}_A \:,
\end{equation}
where $Z_A$ and $\mathbf{R}_A$ are the nuclear charges and positions,
or we evaluate the dipole through its effect on the total energy, 
\begin{equation}
  E(\mathbf{E}) = E(\mathbf{E}=0)  - \mathbf{D} \cdot \mathbf{E}.
\end{equation}
The response to an external static electric field ($\mathbf{E}$) was implemented in MOLGW \cite{bruneval_cpc2016}
for an earlier study about hyperpolarizabilities within RPA \cite{besalu_jctc2023}.
Using two small values for the electric field along the bonding axis of the CO molecule allowed us to estimate the static dipole 
induced by the RPA total energy functional.

\begin{figure}
  \includegraphics[width=0.98\columnwidth]{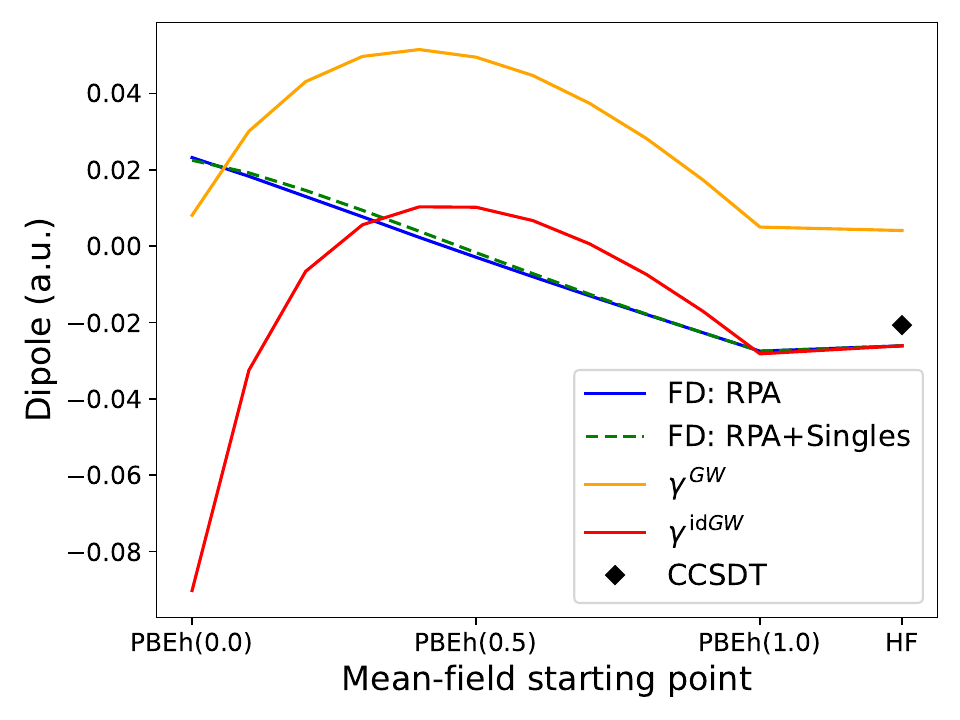}
\caption{
  \label{fig:dipole}
  Static dipole along the CO molecule axis in atomic units within aug-cc-pVTZ basis as a function of the mean-field starting point
  as obtained from finite difference (FD) using RPA total energy, RPA + Singles total energy,
  or from a direct calculation with $\gamma^{GW}$ or $\gamma^{\textrm{id}GW}$ density matrices.
  CCSDT dipole is given as a comparison.
}
\end{figure}

We performed these calculations for several mean-field starting points: HF and a hybrid functional PBEh($\alpha$) where 
the content of exact exchange $\alpha$ has been varied from 0 to 1.
The CO bond length has been fixed to 1.1503~\AA{}.
The results reported in Fig.~\ref{fig:dipole} confirm the comparison with the relaxed density matrices.
The unrelaxed density matrix $\gamma^{GW}$ does not derive from the RPA total energy independent of the mean-field starting point.
The iterated Dyson density matrix $\gamma^{\textrm{id}GW}$ does correspond to the one obtained from the RPA total energy functional for the HF starting point.
When the exact exchange content $\alpha$ is decreased, the discrepancy increases.

For non-HF starting points, the Brillouin theorem no longer holds and a first-order correction to the total energy
should be included.
\cite{ren_prl2011}
This term named ``singles'' was included in Fig.~\ref{fig:dipole} to give the curve labeled RPA+Singles.
The figure shows that the effect of single excitations on the dipole value is very mild and does not 
reconcile the discrepancy between $\gamma^{\textrm{id}GW}$ and the total energies.

We conclude that the iterated Dyson $\gamma^{\textrm{id}GW}$ can be understood as the analytical derivative of a total energy functional only in the case of HF starting point. This reflects the fact that the static part of the $GW$ self-energy is the sum of Hartree and exact exchange potentials, Therefore, the $GW$ self-energy induces relaxation due to the HF Hessian. However, the Z-vector approach constrains the orbital relaxation to minimize the gKS energy upon the addition of the RPA correlation energy. For this reason, the orbitals must relax under the gKS Hessian.

The inclusion of additional terms in the matrix $\boldsymbol{\mathcal{A}}$ to render the density matrix fully relaxed is left for further study.

\section{\label{sec:results} Benchmarking the iterated Dyson density matrix over 34 molecules}

We now turn to the practical evaluation of the density matrices in realistic molecules.
To this end, we use a list of 34 small molecules that were already used in Ref.~\onlinecite{bruneval_jctc2019}, considered in their MP2 optimized geometry and described with spherical Dunning basis sets cc-pVXZ, (X = D,T,Q,5)\cite{woon_jcp1995}. 

\subsection{\label{sec:numverif} Numerical verification}

\begin{figure}[h!]
  \includegraphics[width=0.98\columnwidth]{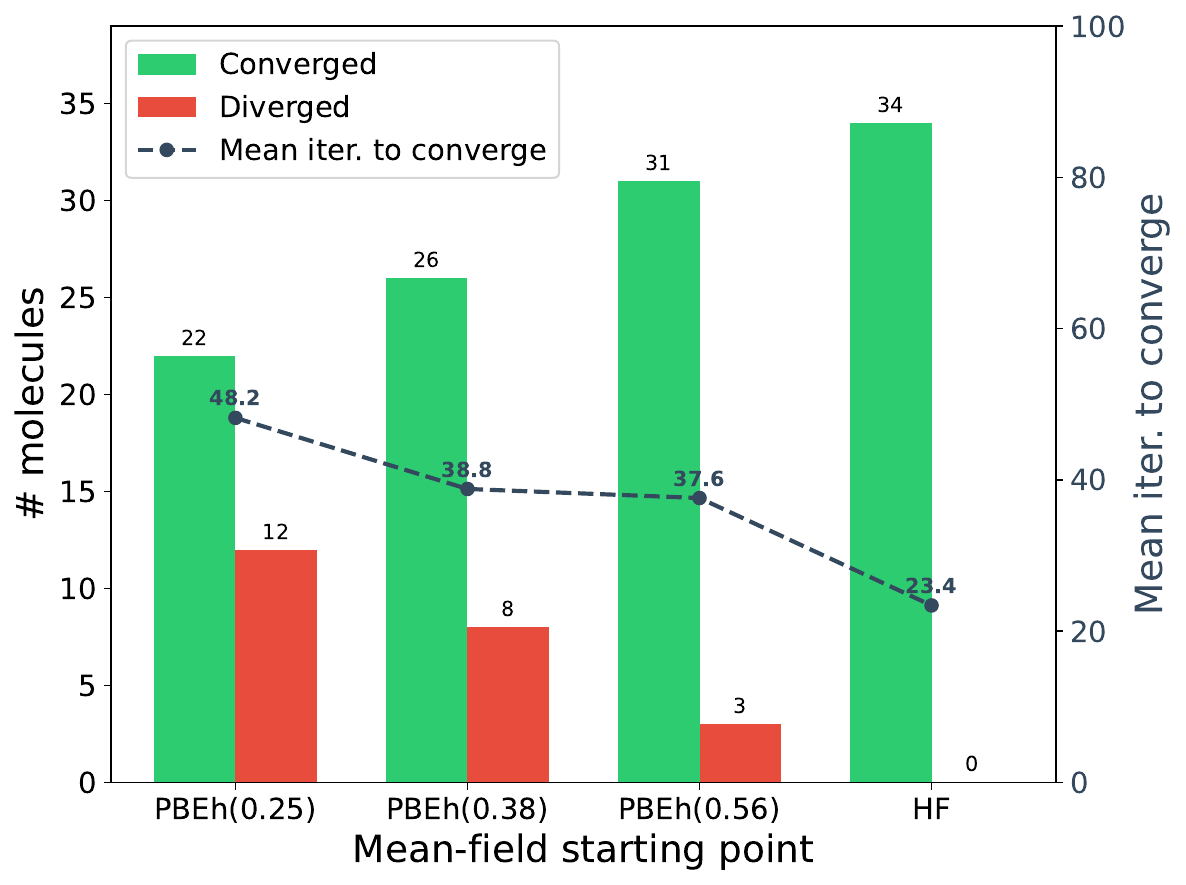}
\caption{
  \label{fig:numver}
  Convergence rate and stability for a benchmark set of 34 small molecules within cc-pVDZ as a function of the mean-field starting point. Convergence criteria: maximum 100 iterations, $10^{-6}$ convergence threshold for each density matrix element.
}
\end{figure}

First, we comment on the different approaches used to obtain the iterated Dyson density matrix, as described in sections~\ref{sec:LinSys} and \ref{sec:NumInt}, that we have implemented in the BAND engine of a modified version of the Amsterdam Modeling Suite (AMS).\cite{Baerends2025} We perform this analysis for the 34 molecules with the cc-pVDZ basis set, using PBE-based hybrid functionals with different fractions of exact exchange as implemented in libXC.\cite{Lehtola2018} Here we focus exclusively on the stability and convergence rate of the numerical approach, using the analytical result obtained from solving the linear system as reference point.

To ensure the correctness of the numerical approach, we calculated the full contribution of the static part of the self-energy to $\Delta\gamma^{\textrm{id}GW}$, rather than restricting the evaluation to the occupied-virtual block. This way, we can observe explicitly that the chosen frequency grid (Gauss-Legendre quadrature with 1024 grid points) correctly yields vanishing occupied-occupied and virtual-virtual blocks. The density matrix was converged using a threshold of $10^{-6}$ per density matrix element, accompanied by a linear mixing scheme where the updated guess for $\Delta\gamma^{\textrm{id}GW}$ consisted of 75\% of the previous guess and 25\% of the newly calculated solution.

In Fig.~\ref{fig:numver}, we report the stability and convergence rate of the 34 molecules, where clear trends emerge for both metrics. With a HF starting point, we consistently converge to the analytical result. However, with a PBEh($\alpha$) starting point, convergence is not guaranteed. Crucially, smaller values of $\alpha$ lead more often to divergence, and even utilizing the analytical result as the initial guess fails to remedy this behavior. Additionally, the HF starting point requires fewer iterations to reach convergence compared to PBEh($\alpha$), with the mean number of iterations needed increasing with $\alpha$. Note that the ratio previous guess and newly calculated solution in the linear mixing affects this number, but this does not affect the observed trend.

To gain insight about the ill-behaving cases, we have performed a systematic diagonalization of Fock Hessian matrices $\boldsymbol{\mathcal{A}}$ in the test set. 
We have observed a clear link between the non-positiveness of $\boldsymbol{\mathcal{A}}$ and non-convergence.
However, this link is likely not strict due to the linear mixing that damps the oscillations.

Together, HF shows to have a well-behaved landscape of $\Delta\gamma^{\textrm{id}GW}$, whereas for PBEh($\alpha$) this landscape becomes more and more ill-behaved for smaller values of $\alpha$.

\subsection{Comparison to coupled-cluster density matrices}

\begin{figure*}[hbt!]
  \includegraphics[width=0.98\textwidth]{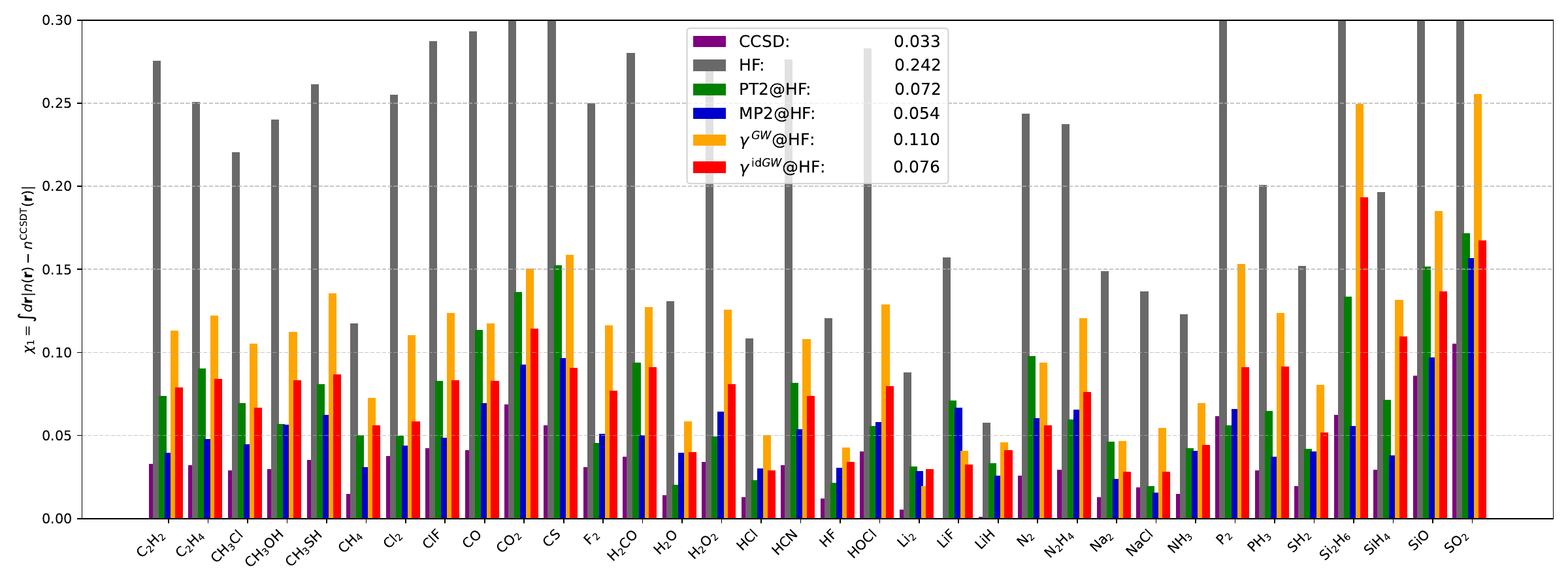}
\caption{
  \label{fig:density34}
 Deviation from coupled-cluster CCSDT for the electronic density 
 as measured by the $\chi_1$ distance
 for a benchmark set of 34 small molecules.
 The mean value for $\chi_1$, averaged across the 34 molecules, is given in the caption.
 PT2 is the second-order perturbation theory density matrix, MP2 is the relaxed version of it.
}
\end{figure*}

\begin{figure}[hbt!]
  \includegraphics[width=0.98\columnwidth]{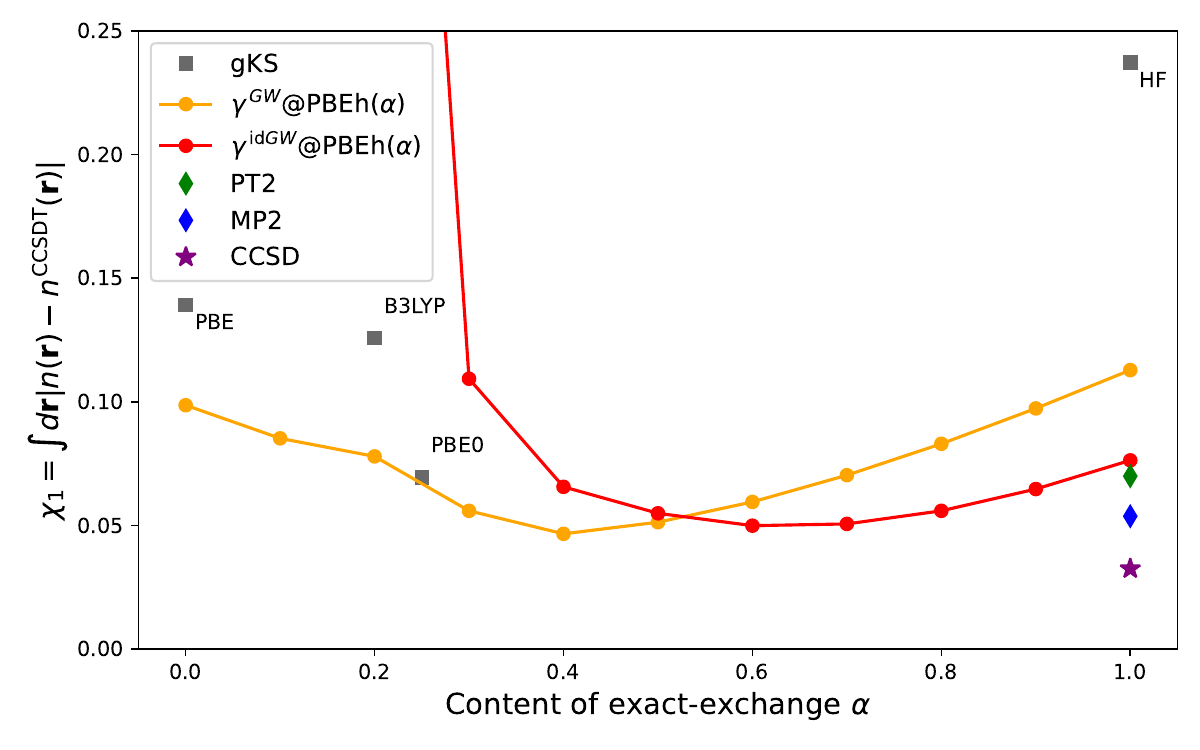}
\caption{
  \label{fig:chi1}
 Deviation from coupled-cluster CCSDT for the electronic density 
 as measured by the $\chi_1$ distance varying the mean-field starting point averaged over the 34 molecules of the set.
}
\end{figure}

In this benchmark, the relaxed coupled-cluster density matrix serves as a reference.
In contrast with the previous study\cite{bruneval_jctc2019} that used CCSD with the frozen core approximation,
here we use the more accurate CCSDT approximation without frozen core.
This was made possible by the use of the CFOUR code. \cite{matthews_jchemphys2020}
Here, the 34 molecules have been considered with the cc-pVQZ basis set.

First, we compare the error in the electronic density induced by the different approximations to the density matrix.
We use the $\chi_1$ measure, which is the unitless distance between a test and reference electronic density:
\begin{equation}
  \chi_1 = \int d \br | n^\mathrm{test}(\br) - n^\mathrm{CCSDT}(\br) | .
\end{equation}
In practice, we evaluate the densities on the accurate spatial grids that are used for the exchange-correlation potential and energy quadratures.
\cite{becke_jcp1988}

In Fig.~\ref{fig:density34}, we report the deviation with respect to the CCSDT electronic density for the 34 molecules.
In this figure, we restrict ourselves to the HF mean field starting point.
We first conclude that $\gamma^{\textrm{id}GW}$ is always an improvement over the original $\gamma^{GW}$ density,
with the only exception of Li$_2$ for which the error is already very small anyway.
For fair comparison purposes, we also include standard approximations: density matrix at second-order perturbation theory
(labeled PT2), and relaxed density matrix at second-order (labeled MP2). These expressions are specified in App.~\ref{app:mp2}
For these molecules, MP2 performs equally or better than $\gamma^{\textrm{id}GW}$.
This is particularly visible for the silicon-containing molecules.

Note that these conclusions deviate from those in Ref.~\onlinecite{bruneval_jctc2019}, in which
$\gamma^{GW}$ was found to to be superior to MP2.
We realize now that the use of the CCSDT reference instead of the CCSD shakes up the ranking of the methods.

\begin{figure*}[hbt!]
  \includegraphics[width=0.98\textwidth]{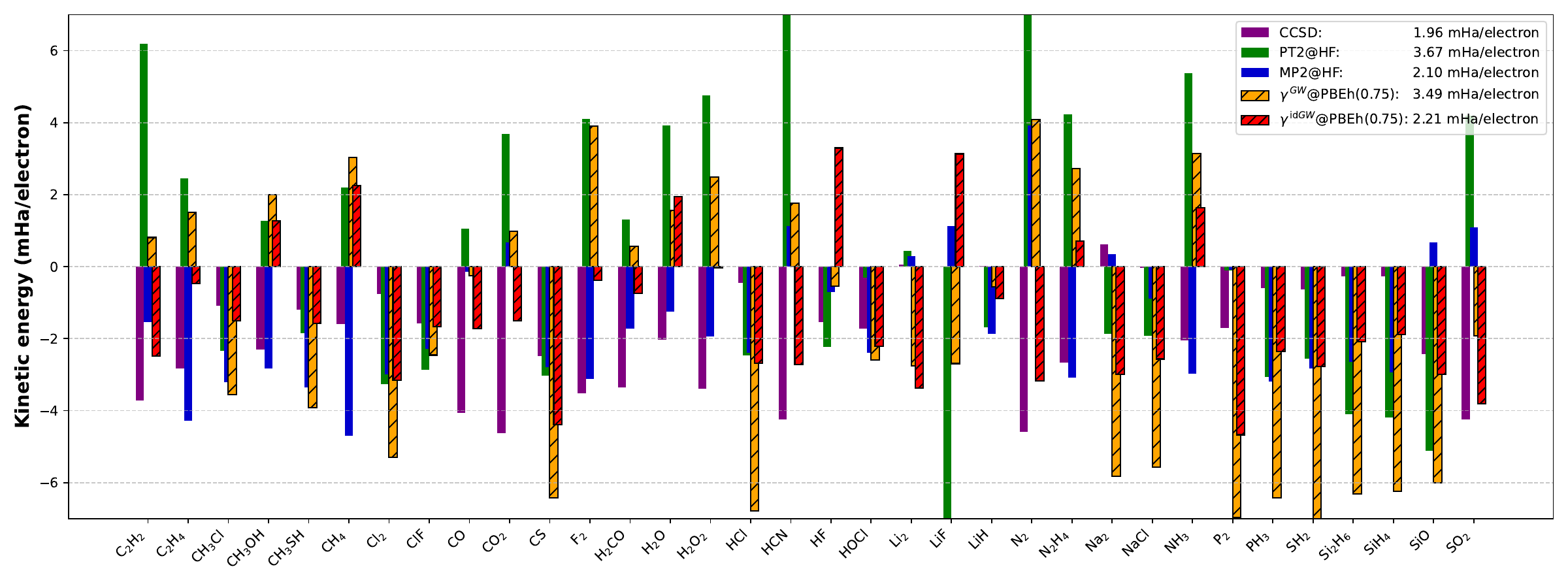}
\caption{
  \label{fig:ek34}
 Deviation against coupled-cluster CCSDT for the kinetic energy
 for a benchmark of 34 small molecules.
 The mean absolute error, averaged across the 34 molecules, is given in the caption.
 Mean-field starting point is PBEh(0.75) for both $\gamma^{GW}$ and $\gamma^{\textrm{id}GW}$.
}
\end{figure*}

Now we monitor the quality of the obtained electronic density when adjusting the starting point in Fig.~\ref{fig:chi1}.
In the figure, we place several generalized Kohn-Sham results for comparison (PBE, B3LYP, PBE0, HF).
We use the PBEh($\alpha$) functional where the content of exact exchange $\alpha$ is varied from 0 to 1.
The quality of the $GW$ densities can be improved with optimal starting points.
For $\gamma^{GW}$, the best $\alpha$ would be 0.4, whereas for $\gamma^{\textrm{id}GW}$ it would be around 0.6-0.7.
With these optimal $\alpha$, the $GW$-based densities are of MP2 quality.
However, it should be noted that low content of exchange is dangerous for $\gamma^{\textrm{id}GW}$,
with errors that skyrocket.
We ascribe this instability to the sensitivity to small HOMO-LUMO gaps.
It is consistent with the difficulty of the numerical
approach to reach a stationary point in Section~\ref{sec:numverif}.

After having tested the electronic density, which is a contraction of the density matrix, we would like to turn to 
the benchmarking of a quantity that is beyond the mere density.
Hence, we consider the kinetic energy itself,
\begin{equation}
  K = -\frac{1}{2} \int d \br \lim_{\br' \to \br} \nabla_{\br'}^2 \gamma(\br, \br') .
\end{equation}
The comparison with respect to CCSDT is reported in Fig.~\ref{fig:ek34}.
In this plot, we did not reproduce the HF kinetic energy, since it is completely missing the correlation part
of the kinetic energy that is very significant.
Again we confirm that with the exception of very few counter examples, $\gamma^{\textrm{id}GW}$ always yields improved kinetic energies
over $\gamma^{GW}$.
It is quite remarkable that MP2 and $\gamma^{\textrm{id}GW}$@PBEh(0.75) have almost the same mean absolute error as CCSD.
Very similar conclusions could have been drawn from the deviation of the exchange energy (not shown here).

\section{\label{sec:conclusion}Conclusion}

In this study, we have analyzed the effect of iterating the static part of the self-energy in the Dyson equation.
This technique was first proposed in the context of ADC by Cederbaum and coworkers \cite{vonniessen_cpr1984}
and yields linear equations that can be solved without the need
to actually iterate the equations.

The possibility to calculate the iterated density matrix with a one-shot calculation
contrasts with the various self-consistent schemes that exist in the literature
which range from fully self-consistent $GW$ (sc$GW$) \cite{Pokhilko2021b},
to quasiparticle self-consistent QS$GW$ \cite{vanschilfgaarde_prl2006,bruneval_springer2014},
passing through ``Fock loop self-consistent''
introduced by Backhouse and coworkers \cite{backhouse_jctc2020,backhouse_jctc2025}.
This last scheme is rather similar to our approach but differs for two main reasons:
they update the molecular orbitals and eigenvalues in the dynamical part of the self-energy
along with their loop cycles;
they use the non-linearized Dyson equation, whereas we use the linearized one.
The complete Dyson equation may look superior however it requires a tuning of the chemical potential $\mu$
in order to conserve the number of electrons.

We showed that iterating the static part of the self-energy in the Dyson equation produces a density matrix that coincides with the one derived from variational procedures in the case 
of a HF starting Green's function.
If applied to the PT2 self-energy (also named GF2), we obtained the so-called relaxed MP2 density matrix.\cite{amos1980corrections, trucks_cpl1988}
This connection is derived in App.~\ref{app:mp2}.
If applied to the $GW$ self-energy, we get the relaxed RPA density matrix.
However, when using a generalized Kohn-Sham Green's function that is not HF, the iterated Dyson density matrix 
deviates from the relaxed one,
mostly due to the use of the HF orbital Hessian instead of the gKS one.

Then we have benchmarked the quality of the obtained density matrix against coupled-cluster (CCSDT) reference values.
We showed that the iterated Dyson $GW$ density matrix $\gamma^{\textrm{id}GW}$ systematically improves upon 
the non-iterated counterpart $\gamma^{GW}$ for mean-field starting points with a large content of exact-exchange.
When using it with a low content of exact-exchange, the iteration procedure can be dangerous and can lead to a divergence of the scheme. This originates from the evaluation
of the HF orbital Hessian but with gKS wavefunctions and eigenvalues.

The iterated Dyson procedure can be viewed as an additional step towards the fully self-consistent evaluation of the Green's function
while keeping the ease of working in the original mean-field MO space.

\appendix

\section{\label{app:alternate} Convenient alternate form for the linear system}

We will focus on calculating the correction with respect $\Delta \gamma_{ia}^{\textrm{id}GW}$ that we define as
\begin{equation}
  \Delta \Delta \gamma_{ia}^{\textrm{id}GW} = \Delta \gamma_{ia}^{\textrm{id}GW} - \Delta \gamma_{ia}^{GW}.
\end{equation}

Introducing this notation in Eq.~(\ref{eq:idgwov}), we obtain
\begin{align}
 \Delta \Delta \gamma_{ia}^{\textrm{id}GW}  = &
\frac{1}{\epsilon_i - \epsilon_a}
          \sum_{jb} \left( \langle  ij || ab \rangle 
                        +  \langle  ib || aj \rangle
                    \right) \nonumber \\
        &           \cdot
                     (\Delta \Delta \gamma_{jb}^{\textrm{id}GW} + \Delta \gamma_{ia}^{GW} ) \nonumber \\
        & + \frac{1}{\epsilon_i - \epsilon_a}
            \sum_{kl} \langle  ik || al \rangle \Delta \gamma_{kl}^{GW}  \nonumber \\
        & + \frac{1}{\epsilon_i - \epsilon_a}
            \sum_{bc} \langle  ib || ac \rangle \Delta \gamma_{bc}^{GW}  
\end{align}

This form allows one to regroup all the terms containing $\Delta \gamma^{GW}$ in the same double sum over all the states:
\begin{equation}
\label{eq:linearsystem2}
 \sum_{jb} \mathcal{A}_{ia, jb} \Delta \Delta \gamma_{ia}^{\textrm{id}GW}  = 
         - \sum_{mn} \langle  im || an \rangle \Delta \gamma_{mn}^{GW} ,
\end{equation}
where matrix elements $\mathcal{A}_{ia, jb}$ are those defined earlier in Eq.~(\ref{eq:linear_system_a}).

After the solution of the linear system in Eq.~(\ref{eq:linearsystem2}),
the final density matrix is obtained as
$\gamma_{ia}^\gKS + \Delta \gamma_{ia}^{GW} + \Delta \Delta \gamma_{ia}^{GW}$.

\section{\label{app:mp2} Application to second-order perturbation theory}

The self-energy at second-order perturbation theory has been known for a long time: \cite{szabo_book}
\begin{align}
  \label{eq:pt2_analytic}
  M_{pq}^\mathrm{PT2}(\omega) = &\frac{1}{2} \sum_{k ia} 
                                            \frac{ \langle p a || k i \rangle \cdot \langle q a || k i\rangle }
                                    {\omega - \epsilon_k + \epsilon_a - \epsilon_i - \mi \eta } \nonumber\\
                 & + \frac{1}{2} \sum_{c ia}
                                            \frac{ \langle p i || c a \rangle \cdot \langle q i || c a\rangle }
                                    {\omega - \epsilon_c - \epsilon_a + \epsilon_i  + \mi \eta} .
\end{align}

Then performing the integral in Eq.~(\ref{eq:deltagamma}) analytically,
we obtain the expressions for the different blocks of the PT2 density matrix:
\begin{subequations}
\begin{align}
  \label{eq:pt2oo}
  \Delta \gamma^\mathrm{PT2}_{ij}
     = & - \frac{1}{2}\sum_{k ac} 
                                 \frac{ \langle i k || c a\rangle }
                                     {\epsilon_i - \epsilon_c - \epsilon_a + \epsilon_k}  \nonumber \\
        & \quad  \cdot 
                                 \frac{ \langle j k || c a\rangle }
                         {\epsilon_j - \epsilon_c - \epsilon_a + \epsilon_k} \\
  \label{eq:pt2vv}
  \Delta \gamma^\mathrm{PT2}_{ab}
     = & \phantom{-}\frac{1}{2} 
                 \sum_{ik c} 
                           \frac{ \langle a c || k i \rangle }
                                {\epsilon_a - \epsilon_k + \epsilon_c - \epsilon_i} \nonumber \\
                     & \quad    \cdot 
                           \frac{ \langle b c || k i \rangle }
                                    {\epsilon_b - \epsilon_k + \epsilon_c - \epsilon_i} \\
  \label{eq:pt2ov}
  \Delta \gamma^\mathrm{PT2}_{ib}
     = & 
       \frac{1}{2} \frac{1}{\epsilon_i - \epsilon_b}
       \Bigg\{
      \sum_{jk a}
                  \frac{ \langle i a || k j \rangle \cdot \langle b a || k j \rangle }
                     {\epsilon_a + \epsilon_b - \epsilon_j - \epsilon_k} \nonumber\\
      & 
       \phantom{ \frac{1}{\epsilon_i - \epsilon_b} }
      + \sum_{jac}
                  \frac{ \langle i j || c a \rangle \cdot \langle b j || c a \rangle }
                     {\epsilon_i + \epsilon_j - \epsilon_a - \epsilon_c}  
                      \nonumber\\
      & 
       \phantom{ \frac{1}{\epsilon_i - \epsilon_b} }
       + \langle i | \Sigma_x[\gamma^\gKS] - v_{xc}[\gamma^\gKS] | b \rangle
       \Bigg\}  .
\end{align}
\end{subequations}

When used to create the Hartree potential and Fock exchange, these blocks are precisely those that induce the static diagrams (A diagrams in Cederbaum's notation) of the 
third-order self-energy. \cite{vonniessen_cpr1984}

Taking  $\Delta \gamma^\mathrm{PT2}_{ib} = 0$ yields an unrelaxed MP2 density matrix \cite{frisch_chemphyslett1990}
that does not couple occupied and virtual states.
The density matrix in Eqs.~(\ref{eq:pt2oo}-\ref{eq:pt2ov}) is the one we use in this work with label PT2.

Then, introducing this expression in the iterated Dyson procedure follows the exact same steps as for the $GW$ density matrix.
We end up solving a linear system $\boldsymbol{\mathcal{A}} \cdot \boldsymbol{\mathcal{X}} = \boldsymbol{\mathcal{Y}}$
as in Sec.~\ref{sec:LinSys}, where
\begin{subequations}
\begin{align}
  \mathcal{A}_{ia, jb} = &  
(\epsilon_a - \epsilon_i) \delta_{ij} \delta_{ab} 
  + \langle  ij || ab \rangle 
      +  \langle  ib || aj \rangle \\
  \label{eq:linear_system_x_pt2}
  \mathcal{X}_{jb} = & \Delta \gamma_{jb}^{\textrm{idPT2}}  \\
  \label{eq:linear_system_y_pt2}
  \mathcal{Y}_{ia} = & -\sum_{kl} \langle  ik || al \rangle \Delta \gamma_{kl}^\mathrm{PT2}   \nonumber  -\sum_{bc} \langle  ib || ac \rangle \Delta \gamma_{bc}^\mathrm{PT2}   \nonumber \\
           & +(\epsilon_a - \epsilon_i)  \Delta \gamma_{ia}^\mathrm{PT2} .
\end{align}
\end{subequations}

This linear system coincides precisely with the relaxed MP2 density matrix equation \cite{frisch_chemphyslett1990, amos1980corrections, trucks_cpl1988}
if we select a HF starting point, which makes the last term of Eq.~(\ref{eq:pt2ov}) equal to zero.

\begin{acknowledgments}
This work was performed using HPC resources from GENCI-CCRT-TGCC (Grants No. 2025-096018). AF acknowledges funding through a VENI grant from NWO under grant agreement VI.Veni.232.013.
\end{acknowledgments}

\section*{Author contibution}
\textbf{Fabien Bruneval:} Conceptualization (lead), Methodology (lead), Software (equal), Validation (equal), Investigation (equal), Writing - Original Draft (lead), Writing - Review \& Editing (equal), Visualization (equal).
\textbf{Erik Verzijl:} Methodology (supporting), Software (equal), Validation (equal), Investigation (supporting), Writing - Original Draft (supporting), Writing - Review \& Editing (equal), Visualization (supporting).
\textbf{Arno Förster:} Writing - Review \& Editing (equal), Supervision (equal), Funding acquisition (equal). \textbf{Mauricio Rodriguez-Mayorga:} Writing - Review \& Editing (equal), Software (supporting), Investigation (supporting).

\section*{Data Availability Statement}
The data, scripts, figures that support the findings of this study are publicly and permanently hosted in Zenodo \cite{zenodo_supplmat}.

%

\end{document}